\documentclass[11pt]{article}
\usepackage{amsmath}
\usepackage{amsfonts}
\def\p{\partial}
\def\be{\begin{equation}}
\def\ee{\end{equation}}
\def\bea{\begin{eqnarray}}
\def\eea{\end{eqnarray}}

\def\scri{\mathcal{I}}

\begin{document}

\title{Some geometry of de Sitter space}
\author{Paul Tod}

\maketitle
\begin{abstract}
In this expository note, I present some basic geometric and twistor theoretic facts about de Sitter space leading up to a discussion of Penrose's quasi-local mass construction for linear gravity 
theory in the de Sitter background. The corresponding results for Minkowski space and anti--de Sitter space are familiar and can be found in \cite{PR} or \cite{HT} for the former and \cite{KT} for the latter. 
Some of the formulas given here are also quite familiar, but some are thought to be new and it is convenient to have these facts collected in one place.
\end{abstract}

\section{The metric of de Sitter space}
\begin{enumerate}
\item
The simplest definition of 4-dimensional de Sitter space is as the hyperboloid
\be\label{d1}U^2-X^2-Y^2-Z^2-W^2=-H^{-2}\ee
for $H>0$, in the 5-dimensional Minkowski space with metric
\[ds^2=dU^2-dX^2-dY^2-dZ^2-dW^2.\]
In this form it is clear that the isometry group is the pseudo-orthogonal group $O(1,5)$.
\item The whole manifold can be covered by the coordinatisation
\begin{eqnarray*}U=H^{-1}\sinh Ht,\;&&X=H^{-1}\cosh Ht\sin r\sin\theta\cos\phi,\\
Y=H^{-1}\cosh Ht\sin r\sin\theta\sin\phi,\;&&Z=H^{-1}\cosh Ht\sin r\cos\theta,\\
W=H^{-1}\cosh Ht\cos r,
\end{eqnarray*}
when the metric is
\be\label{d4}
g=dt^2-H^{-2}\cosh^2Ht(dr^2+\sin^2r(d\theta^2+\sin^2\theta d\phi^2)).\ee

As is well-known (and will be shown below), the Weyl tensor of this metric is zero and the Ricci tensor is proportional to the metric. With conventions as in \cite{PR} we have
\begin{eqnarray*}
R_{abcd}&=&H^2(g_{ac}g_{bd}-g_{ad}g_{bc})\\
R_{ab}&=&3H^2g_{ab}\nonumber
\end{eqnarray*}
so that the Ricci scalar is $R=12H^2$ and the cosmological constant is $\lambda=3H^2$.
\item
It will be convenient to write the metric (\ref{d4}) in the form
\be\label{d14}
ds^2=dt^2-H^{-2}\cosh^2 Ht\,h_{ij}dx^idx^j
\ee
and then $h_{ij}$ is the metric on the unit $S^3$ in flat $\mathbb{R}^4$. This metric is Einstein with scalar curvature $s_h= 6$.

\item To add the conformal infinity of de Sitter space as a boundary, first introduce conformal time $\tau$ by
\[d\tau=\mbox{sech }Htdt,\]
so that w.l.o.g.
\[\cosh Ht=\mbox{csc}H\tau,\;\;\sinh Ht=-\mbox{cot}H\tau;\]
now the range $-\infty<t<\infty$ in proper time is $0<\tau<\pi/H$ in conformal time, and the conformal boundary is at $\tau=0$ in the past (call this $\scri^-$) and $\tau=\pi/H$ in the future (or $\scri^+$).

\item To add the conformal boundary in a more geometrical way recall the definition of compactified Minkowski space as the projective $O(2,4)$ null cone, equivalently
as the space of generators of the $O(2,4)$ null cone
\be\label{d5}V^2+U^2-X^2-Y^2-Z^2-W^2=0,\ee
when the metric is
\[
ds^2=dV^2+dU^2-dX^2-dY^2-dZ^2-dW^2.\]
Evidently the hyperboloid (\ref{d1}) is the slice $V=H^{-1}$ of the null cone (\ref{d5}). In a similar fashion, Minkowski space is realised as the slice $V+W=1$ and
anti--de Sitter space as the (universal cover of the) slice $W=H^{-1}$.

The slice $V=H^{-1}$ misses those generators of the null cone (\ref{d5}) with $V=0$ and these constitute two copies of $S^3$, with $U>0$ and $U<0$ respectively. These are $\scri^+$ and $\scri^-$ respectively.
\item
 We may find the Killing vectors of de Sitter space by decomposing an arbitrary Killing vector as \[K=A\partial_t+B^i\partial_i.\] Then the Killing vector equation can readily be solved to find
\be\label{kv1}
K=F(x_i)\partial_t+H\tanh H t\,h^{ij}F_{,i} \partial_j+X^i(x^j)\partial_i,
\ee
where $h_{ij}$ is as in (\ref{d14}), $X^i\partial_i$ is a Killing vector for $h_{ij}$, and $F$ is a \emph{conformal scalar} for $h_{ij}$, by which we mean
 \be\label{cs1}D_iD_jF=-Fh_{ij},\ee
 where $D_i$ is the Levi-Civita derivative for $h_{ij}$. The space of such $F$ is 4-dimensional, since they are induced by linear functions of the Cartesian coordinates in the ambient flat $\mathbb{R}^4$, and the space of such $X^i$ is 6-dimensional,
giving the 10-dimensional space of Killing vectors for de Sitter space.

By inspection (though it is a general result) any Killing vector $K$ extends to $\scri$, and there it reduces to
\[h^{ij}F_{,i} \partial_j+X^i(x^j)\partial_i,\]
and is a conformal Killing vector of $h_{ij}$.

We shall consider the conformal Killing vectors of de Sitter space in section 3 below.

\end{enumerate}
\section{Twistors in de Sitter space}
\begin{enumerate}
\item In de Sitter space the (four-dimensional) twistor equations  are
\be\label{t1}\nabla_{AA'}\omega_B=-i\epsilon_{AB}\pi_{A'},\;\;\nabla_{AA'}\pi_{B'}=-i\Lambda\epsilon_{A'B'}\omega_{A},\ee
where $\Lambda=R/24=H^2/2$ is the curvature scalar in Newman-Penrose conventions. These are integrable in the sense of having a 4-complex dimensional vector space of solutions, \emph{the twistor space} $\mathbb{T}$. As is conventional, we write the elements of $\mathbb{T}$
as \emph{twistors} $X^\alpha, Y^\alpha, Z^\alpha...$ where each element corresponds to a pair of spinor fields $(\omega^A,\pi_{A'})$.
\item As with Minkowski space, the conformal group acts on $\mathbb{T}$, in that the Lie derivative along any conformal Killing vector of spinor fields $(\omega^A(x),\pi_{A'}(x))$ satisfying (\ref{t1}) also
satisfies (\ref{t1}).
\item From (\ref{t1}) we see that $\Sigma(Z,\hat{Z})$ defined for two linearly independent twistors $Z^\alpha, \hat{Z}^\alpha$, with corresponding
spinor fields $(\omega^A,\pi_{A'}), (\hat{\omega}^A,\hat{\pi}_{A'})$,  by
\be\label{t2}
\Sigma=\omega^A\bar{\hat{\pi}}_{A}+\pi_{A'}\bar{\hat{\omega}}^{A'}
\ee
is constant (as a scalar field) and so defines a (pseudo) Hermitian form on $\mathbb{T}$
which we may write as $\Sigma_{\alpha\alpha'}Z^\alpha\bar{\hat{Z}}^{\alpha'}$. It is a standard piece of twistor theory that this form has signature $(++--)$ and so
in particular it is nondegenerate and therefore invertible. Write the inverse as $\Sigma^{\alpha\alpha'}$. This Hermitian form is conventionally
used to eliminate primed twistor indices, so that for example $Y^{\alpha'},W_{\beta'}$ are written in terms of $Y_\alpha:=\Sigma_{\alpha'\alpha}Y^{\alpha'}, W^\beta:=\Sigma^{\beta'\beta}W_{\beta'}$. With this
usage, complex conjugation is viewed as a map to the dual of $\mathbb{T}$ rather than to the complex conjugate of $\mathbb{T}$:
\[Z^\alpha\rightarrow \bar{Z}^{\alpha'}\rightarrow \bar{Z}_\alpha=\Sigma_{\alpha\alpha'}\bar{Z}^{\alpha'}.\]

\item There is also a canonical 4-form on $\mathbb{T}$ which can be defined on a choice of four twistors $Z^\alpha_1,...,Z^\alpha_4$ by choosing a normalised spinor dyad (and its complex conjugate) and forming the determinant whose rows are the four functions $\omega^0,\omega^1,\pi_{0'},\pi_{1'}$ which are components in the dyads for each twistor. This determinant is constant
as a scalar field by (\ref{t1}) and so determines a 4-form as det$=\epsilon_{\alpha\beta\gamma\delta}Z^\alpha_1...Z^\delta_4$. Both $\Sigma$ and $\epsilon$ are conformally invariant, and
the endomorphisms of $\mathbb{T}$ preserving $\Sigma$ and $\epsilon$ make up the group $SU(2,2)$.

\item One more canonical structure is given on $\mathbb{T}$ in de Sitter spacee. This is the anti-linear map
\[N:(\omega^A,\pi_{A'})\rightarrow (\bar{\pi}^A,-\Lambda\bar{\omega}_{A'})\]
which can be seen to map solutions of (\ref{t1}) to solutions. It can be written as a map on $\mathbb{T}$
as
\be\label{t3}Z^\alpha\rightarrow\hat{Z}^\alpha=N^\alpha_{\beta'}\bar{Z}^{\beta'}.\ee
There will be an adjoint $N^{\alpha'}_\beta$ and then it can be seen that
\begin{eqnarray}
N^\alpha_{\beta'}N^{\beta'}_\gamma&=&-\Lambda\delta^\alpha_\gamma\\
\Sigma_{\alpha\alpha'}N^\alpha_{\beta'}N^{\alpha'}_{\beta}&=&\Lambda\Sigma_{\beta\beta'}\\
\Sigma_{\alpha\alpha'}N^{\alpha'}_{\beta}&=&-\Sigma_{\beta\alpha'}N^{\alpha'}_{\alpha}.
\end{eqnarray}
\item Motivated by the last of these we can introduce $I_{\alpha\beta}:=\Sigma_{\alpha\beta'}N^{\beta'}_\beta=-I_{\beta\alpha}$ which is the \emph{infinity twistor} for de Sitter space. In
coordinates with $Z^\alpha=(\omega^A,\pi_{A'})$ we have $I_{\alpha\beta}=\mbox{diag}(\Lambda\epsilon_{AB},\epsilon^{A'B'})$. The subgroup of $SU(2,2)$ preserving the infinity twistor will 
be (a finite cover of) the de Sitter group.

The infinity twistor satisfies various identities including
\[\frac12\epsilon^{\alpha\beta\gamma\delta}I_{\gamma\delta}=\bar{I}^{\alpha\beta},\;\;I_{\alpha\beta}\bar{I}^{\beta\gamma}=-\Lambda\delta_\alpha^\gamma.\]
The overbar on $I^{\alpha\beta}$ is therefore redundant and can be omitted.
\end{enumerate}
\section{Killing and conformal Killing vectors in de Sitter space}
\begin{enumerate}
\item Killing and conformal Killing vectors can be constructed from twistors as follows: suppose that $Z^\alpha, \hat{Z}^\alpha$ are two linearly independent twistors with corresponding
spinor fields $(\omega^A,\pi_{A'}), (\hat{\omega}^A,\hat{\pi}_{A'})$  and define vector fields
\begin{eqnarray*}
K^a&=&\omega^A\hat{\pi}^{A'}+\hat{\omega}^A\pi^{A'}\\
L^a&=&\hat{\omega}^A\pi^{A'}-\omega^A\hat{\pi}^{A'}\\
\tilde{K}^a&=&\bar{\pi}^A\pi^{A'}-\Lambda\omega^A\bar{\omega}^{A'}\\
\tilde{L}^a&=&\bar{\pi}^A\pi^{A'}+\Lambda\omega^A\bar{\omega}^{A'}
\end{eqnarray*}
then, by virtue of (\ref{t1}), $K^a$ and $\tilde{K}^a$ are Killing vectors, and $L^a$ and $\tilde{L}^a$ are conformal Killing vectors. Note that $K^a$ and $L^a$ are complex in general, that $\tilde{K}^a$ and $\tilde{L}^a$ are real, and that $\tilde{K}^a$ is space-like and $\tilde{L}^a$ is time-like and future-pointing.

We claim that $L^a$ and $\tilde{L}^a$ are also gradients. To see this introduce
\be\label{k1}
\phi=\Lambda\omega_A\hat{\omega}^A-\pi_{A'}\hat{\pi}^{A'}
\ee
and calculate
\be\label{k2}
\nabla_a\phi=2i\Lambda L_a
\ee
so that $L_a$ is indeed a gradient, while with
\be\label{k3}
\psi=-i(\omega_A\bar{\pi}^A-\bar{\omega}_{A'}\pi^{A'}),
\ee
which is real, calculate
\be\label{k4}
\nabla_a\psi=2\tilde{L}_a
\ee
and so $\tilde{L}_a$ is also a gradient.

\item We noted above that $\tilde{L}^a$ is a conformal Killing vector which is time-like and future-pointing. For future use we note that, in de Sitter space, there are no Killing vectors which are everywhere time-like and future-pointing. One way to see this is to recall that Killing vectors, for which we have found explicit expressions, extend to $\scri$ and are then tangent to $\scri$; since $\scri$ is space-like the Killing vectors must be space-like or zero at $\scri$ and so cannot be time-like throughout a neighbourhood of $\scri$. Another way is to consider a general Killing vector $X^a$ at a surface $\Sigma$ of constant $t$ in the metric
    (\ref{d4}). If $S$ is a 2-surface on $\Sigma$ bounding a ball $B$ on $\Sigma$ then we have the identity
    \[\oint_S\nabla_aX_bdS^{ab}=\int_B\Box X_ad\Sigma^a=6\Lambda\int_BX_aN^ad\Sigma,\]
where $N^a$ is the unit normal, time-like and future-pointing, $\Box=\nabla_c\nabla^c$, and we use the Killing vector identity
\[\Box X_a=6\Lambda X_a.\]
As $B$ expands to sweep out $\Sigma$ we arrive at
\[\int_\Sigma X_aN^ad\Sigma=0,\]
which shows that $X^a$ cannot be time-like and future-pointing everywhere on $\Sigma$.

\item We can go further with (\ref{k2}) and calculate
\[\nabla_a\nabla_b\phi=2i\Lambda\nabla_aL_b=2\Lambda\phi g_{ab},\]
making use of (\ref{t1}). As in Section 1.6, we call any solution $\chi$ of
\be\label{k5}
\nabla_a\nabla_b\chi=2\Lambda\chi g_{ab}
\ee
a \emph{conformal scalar} then $\phi$ from (\ref{k1}) is a complex conformal scalar and $\psi$ from (\ref{k3}) is a real one. It is a familiar fact that, given an umbilic surface in flat space,
linear functions of the Cartesian coordinates on the ambient flat space restrict to conformal scalars on the surface (in the induced metric). Thus here linear functions in $(U,X,Y,Z,W)$
define conformal scalars and so there is a 5-dimensional vector space of them.

\item It may be less familiar but is easy to see that the 15-dimensional vector space of conformal Killing vectors can be expressed as a sum of a 10-dimensional vector space of Killing vectors and a 5-dimensional vector space of gradient conformal Killing vectors. To see this, suppose $X^a$ is any conformal Killing vector so that
    \[\nabla_aX_b=M_{ab}+Qg_{ab},\]
    for some $Q$ and $M_{ab}$ with $M_{ab}=-M_{ba}.$ Commute derivatives on $X^a$ and rearrange to find
    \[\nabla_cM_{ab}=(Q_a-2\Lambda X_a)g_{bc}-(Q_b-2\Lambda X_b)g_{ac}\]
    where $Q_a=\nabla_aQ$. Commute derivatives on $M_{bc}$ to find
    \[\nabla_a\nabla_bQ=2\Lambda Qg_{ab},\]
so that $Q$ is a conformal scalar. Now it is apparent that $K^a:=X^a-Q^a/2\Lambda$ is a Killing vector so that $X^a$ has been written as a sum of a Killing vector and a gradient
conformal Killing vector.

\item To obtain the whole 5-dimensional space of conformal scalars, note that $\phi=\phi(Z,\hat{Z})$ in (\ref{k1}), regarded as a function on $\mathbb{T}$, is antisymmetric. There is a 6-dimensional
vector space of antisymmetric 2-tensors on $\mathbb{T}$ which can be written $Q_{\alpha\beta}=-Q_{\beta\alpha}$. However the particular antisymmetric combination
\[I_{\alpha\beta}Z^\alpha\hat{Z}^\beta=\Lambda\omega_A\hat{\omega}^A+\pi_{A'}\hat{\pi}^{A'}\]
gives a constant on space-time. Now the 5-dimensional vector space of $Q_{\alpha\beta}$ modulo $I_{\alpha\beta}$ gives the conformal scalars. These can be characterised as trace-free
$Q_{\alpha\beta}$ in the sense that $Q_{\alpha\beta}I^{\alpha\beta}=0$.

The real conformal scalars, like $\psi$ in (\ref{k3}), are now of the form
\[\psi=Q_{\alpha\beta}Z^\alpha I^{\beta\gamma}\bar{Z}_\gamma\]
provided $Q_{\alpha\beta}I^{\beta\gamma}=I_{\alpha\beta}\bar{Q}^{\beta\gamma}$. There is a 5-real dimensional vector space of trace-free such $Q$.

\item The gradient conformal Killing vectors have an interesting form at $\scri$. Set $H=1$ and take the conformal scalar to be induced by the linear expression
\[\psi=\alpha U+\beta X+\gamma Y+\delta Z+\epsilon W\]so that, in the coordinates of section 1, 
\[\psi=\alpha\sinh t+\cosh t(\beta\sin r\sin \theta\cos\phi+\gamma\sin r\sin\theta\sin\phi+\delta\sin r\cos\theta+\epsilon\cos r).\]
Then
\[L^a\partial_a=g^{ab}\phi_a\partial_b=\mbox{sech}\; t\frac{\partial\psi}{\partial t}\partial_\tau-(\mbox{sech}^2\;t)h^{ij} \frac{\partial\psi}{\partial x^i}\partial_{x^j}\]
when the spatial part of the de Sitter metric (\ref{d4}) is $-\cosh^2t \;h_{ij}$. At $\scri$ this reduces to
\be\label{k6}L^a\partial_a=N(r,\theta,\phi)\partial_\tau,
\ee
with $N=\alpha+\beta\sin r\sin \theta\cos\phi+\gamma\sin r\sin\theta\sin\phi+\delta\sin r\cos\theta+\epsilon\cos r$. Thus $L$ extends to $\scri$ in the rescaled metric and is a function $N$ times the unit normal to $\scri$, where $N$ is a constant plus a conformal scalar for the standard $S^3$-metric on $\scri$.

In particular, the gradient conformal Killing vectors are all time-like at $\scri$ (except where they vanish).
\item Symmetrised products of the form $\phi^{AB}=\omega^{(A}\hat{\omega}^{B)}$ define \emph{Killing spinors}, since
\[\nabla_{A'(A}\phi_{BC)}=\nabla_{A'(A}(\omega_B\hat{\omega}_{C)})=0\]
by (\ref{t1}), and this is the defining property of a Killing spinor. In fact
\be\label{t10}\nabla_{AA'}\phi_{BC}=-i\epsilon_{A(B}K_{C)A'},\ee
with $K^a$ as in subsection 3.1 above.


\end{enumerate}
\section{Explicit formulas}
If desired, it is possible to obtain explicit expressions for the solutions of (\ref{t1}).
\begin{enumerate}

\item For simplicity set $H=1$ in (\ref{d1}) and use the NP formalism with the tetrad
\[D=\frac{1}{\sqrt{2}}(\partial_t+\mbox{sech}\, t\partial_r),\;\;\Delta=\frac{1}{\sqrt{2}}(\partial_t-\mbox{sech}\, t\partial_r),\;\;\delta=\frac{\mbox{sech}\,t}{\sqrt{2}\sin r}(\partial_\theta-\frac{i}{\sin\theta}\partial_\phi).
\]

This tetrad has spin coefficients as follows:
\[\tau=\kappa=\pi=\nu=\sigma=\lambda=0=\psi_i=\phi_{ij},\]
\[\gamma=-\epsilon=-\frac{1}{2\sqrt{2}}\mbox{tanh}\, t\]
\[\alpha=-\beta=-\frac{1}{2\sqrt{2}}\frac{\cot\theta}{\cosh t\sin r}\]
\[\rho=-\frac{1}{\sqrt{2}}(\mbox{tanh}\, t+\cot r\mbox{sech}\, t),\;\;\mu=  \frac{1}{\sqrt{2}}(\mbox{tanh}\, t-\cot r\mbox{sech}\, t)\]

and the only nonzero curvature component is $\Lambda=1/2$. The Ricci tensor is
 \[R_{ab}=3g_{ab},\]
 the Ricci scalar is $R=12$ and the cosmological constant $\lambda$ is 3. We can restore $H$ as a parameter in the metric and then $\Lambda=R/24=H^2/2=\lambda/6$. (This confirms the expression for the Riemann tensor in section 1.2.)
\item
Now take components of (\ref{t1}) in the implied spinor dyad. The equations separate and the angular parts are solved in terms of the lowest spin-weighted spherical harmonics. The radial/time dependence is simpler in terms of advanced and retarded null coordinates:
\[u:=\tau-r,\;\;v:=\tau+r.\]
The solutions can be given in terms of four complex parameters $(a_1,a_2,b_1,b_2)$, which are coordinates on $\mathbb{T}$, by
\[\omega^0=(\cosh t)^{1/2}\Omega^0,\]
\[\omega^1=(\cosh t)^{1/2}\Omega^1,\]
\[\pi_{0'}=\frac{i}{\sqrt{2}}(\cosh t)^{1/2}P_{0'},\]
\[\pi_{1'}=\frac{i}{\sqrt{2}}(\cosh t)^{1/2}P_{1'},\]
with
\[\Omega^0=\left(\sin(\theta/2)e^{i\phi/2}(a_1\cos(v/2)+a_2\sin(v/2))+\cos(\theta/2)e^{-i\phi/2}(b_1\cos(v/2)+b_2\sin(v/2))\right),\]
\[\Omega^1=\left(\cos(\theta/2)e^{i\phi/2}(a_1\cos(u/2)+a_2\sin(u/2))-\sin(\theta/2)e^{-i\phi/2}(b_1\cos(u/2)+b_2\sin(u/2))\right),\]
\[P_{0'}=\left(\sin(\theta/2)e^{i\phi/2}(a_1\cos(u/2)-a_2\sin(u/2))+\cos(\theta/2)e^{-i\phi/2}(b_1\cos(u/2)-b_2\sin(u/2))\right),\]
\[P_{1'}=\left(\cos(\theta/2)e^{i\phi/2}(a_1\cos(v/2)-a_2\sin(v/2))-\sin(\theta/2)e^{-i\phi/2}(b_1\cos(v/2)-b_2\sin(v/2))\right).\]
Note that all components of $(\omega^A,\pi_{A'})$ blow up on approach to $\scri$, and at the same rate.

\end{enumerate}
\section{Conserved quantities}
Killing and conformal Killing vectors allow the definition of conserved currents:
\begin{enumerate}
\item
Given a conserved energy-momentum tensor $T_{ab}$ as a field on de Sitter space one can make conserved currents $J_a=T_{ab}K^b$ with any Killing vector $K^a$ and also, if $T_{ab}$ is trace-free, as $J_a=T_{ab}L^b$ with any conformal Killing vector $L^a$. Now given any space-like two-surface $S$ spanned by a space-like 3-surface $\Sigma$ one may associate charges with $S$ by integrating over $\Sigma$:
\be\label{c1}Q[S,K]=\int_\Sigma T_{ab}K^bd\Sigma^a,\ee
and there are similar charges made with $L^a$, in the trace-free case.
\item
In the case when $S$ approaches or even lies on $\scri$, convergence of these charges will require fall-off conditions on $T_{ab}$. Simple power-counting suggests that finiteness for $K^a$ requires 
that $(\cosh Ht)^3T_{ab}$ be bounded as $t\rightarrow \infty$, while for $L^a$ one needs $(\cosh Ht)^4T_{ab}$ bounded.

\end{enumerate}
\section{Linear theory in de Sitter space}
Penrose's quasi-local mass construction \cite{PR} is motivated by using twistor theory to convert the 3-surface integrals for conserved charges (\ref{c1}) to 2-surface integrals over
$\partial\Sigma$ when the conserved energy-momentum tensor $T_{ab}$ is the source of a linearised gravitational field. To write out this construction we need to review the linearisation of general relativity about de Sitter space.
\begin{enumerate}
\item Suppose we perturb the de Sitter metric:
\[g_{ab}\rightarrow g_{ab}+\delta g_{ab},\]
and write $h_{ab}:=\delta g_{ab}$. We raise and lower indices with the de Sitter metric and then the perturbation of the contravariant metric is
\[\delta g^{ab}=-h^{ab},\]
and the perturbation in the Christoffel symbols is
\be\label{g0}\delta\Gamma^a_{\;bc}=\frac12(\nabla_bh^a_{\;c}+\nabla_ch^a_{\;b}-\nabla^ah_{bc} ).\ee
As is entirely familiar, the gauge freedom in $h_{ab}$ is
\[h_{ab}\rightarrow h_{ab}+\nabla_aX_b+\nabla_bX_a,\]
for any suitably differentiable vector field, and this freedom can be used to impose the gauge condition
\be\label{g1}
Q^b:=\nabla_a(h^{ab}-\frac12hg^{ab})=0,\ee
where $h=h_{ab}g^{ab}$, by solving
\[\Box X^a-3H^2 X^a+Q^a=0.\]
\item
The perturbation in the Riemann tensor requires care as the background Riemann tensor is not zero, so that the position of indices is significant. The
variation in $R_{abc}^{\;\;\;\;\;d}$ is
\[\delta R_{abc}^{\;\;\;\;\;d}=\nabla_{[a}\nabla_{|c|}h_{b]}^{\;d}-\nabla_{[a}\nabla^dh_{b]c}+H^2(h_{[a}^{\;d}g_{b]c}+h_{c[a}\delta_{b]}^{\;d}).   \]
We can take the trace of this to obtain
\[\delta R_{ab}=\frac12\Box h_{ab}+H^2(4h_{ab}-hg_{ab})\]
with the aid of (\ref{g1}), so that also
\be\label{v5}
\delta R=\delta(g^{ab}R_{ab})=\frac12\Box h-3H^2h,\ee
and then
\[\delta G_a^{\;b}=\delta(g^{bc}R_{ac}-\frac12Rg^b_a)=\frac12\Box(h_a^{\;b}-\frac12 h\delta_a^{\;b})+H^2(h_a^{\;b}+\frac12h\delta_a^{\;b}).\]
Now given a symmetric, divergence-free tensor $T_{ab}$ the linearised Einstein Field Equations are
\be\label{v8}\delta G_a^{\;b}=-\kappa T_a^{\;b},\ee
with $\kappa=8\pi G/c^2$,  and one checks that both sides are divergence-free, as is necessary for consistency.

We shall want to consider the trace-free Ricci tensor, $S_{ab}=R_{ab}-\frac14 Rg_{ab}$. Unperturbed this is zero and the variation is found to be

\be\label{v4}\delta S_{ab}=\frac12\Box(h_{ab}-\frac14 hg_{ab})+H^2(h_{ab}-\frac14 hg_{ab}).\ee

It is convenient to introduce the tensor
\[t_{abcd}:=\delta R_{abcd}=\delta(g_{de}R_{abc}^{\;\;\;\;\;e}),\]
so that
\be\label{t12}
t_{abcd}=\nabla_{[a}\nabla_{|c|}h_{b]d}-\nabla_{[a}\nabla_{|d|}h_{b]c}+\frac12H^2(g_{ac}h_{bd}-g_{bc}h_{ad}+g_{bd}h_{ac}-g_{ad}h_{bc})
\ee
which is easily seen to have Riemann tensor symmetries:
\[t_{abcd}=-t_{bacd}=-t_{abdc},\;\;t_{[abc]d}=0.\]
It can be decomposed into irreducible pieces as
\[t_{abcd}=\delta C_{abcd}+\frac12(g_{ac}\delta S_{bd}+g_{bd}\delta S_{ac}-g_{ab}\delta S_{cd}-g_{cd}\delta S_{ab})\]
\[+\frac{1}{12}\delta R(g_{ac}g_{bd}-g_{bc}g_{ad})+H^2(g_{ac}h_{bd}+g_{bd}h_{ac}-g_{bc}h_{ad}-g_{ad}h_{bc})\]
where $\delta C_{abcd}$ is the variation in the Weyl tensor, $\delta S_{ab}$ is as in (\ref{v4}) and $\delta R$ is as in (\ref{v5}).

We introduce spinor equivalents
\begin{eqnarray*}
\delta C_{abcd}&=&\psi_{ABCD}\epsilon_{A'B'}\epsilon_{C'D'}+\mbox{ H.c.}\\
\delta S_{ab}&=&-2\phi_{ABA'B'}
\end{eqnarray*}
where H.c. stands for Hermition conjugate. We choose not to include $\delta$ on the right in these definitions since the corresponding Weyl and Ricci spinors are zero in the background. In terms of these spinor fields the Bianchi identities take the usual form:
\bea\label{k9}
\nabla^D_{A'}\psi_{ABCD}&=&\nabla^{B'}_{(A}\phi_{BC)A'B'},\\\nonumber
\nabla^{BB'}\phi_{ABA'B'}&=&-\frac18\nabla_{AA'}\delta R.
\eea
\item
Given the perturbations in curvature and a choice of two twistors $\omega^A,\hat{\omega}^B$ we construct the two-form
\be\label{v6}
f_{ab}=i\psi_{ABCD}\omega^C\hat{\omega}^D\epsilon_{A'B'}-i\phi_{A'B'CD}\omega^C\hat{\omega}^D\epsilon_{AB}+\frac{i}{12}\delta R\omega_{(A}\hat{\omega}_{B)}\epsilon_{A'B'},\ee
and then calculate
\[\nabla^bf_{ab}=\frac{\kappa}{2}T_{ab}K^b.\]
Here we use the linearised Einstein equation (\ref{v8}), the Bianchi identities (\ref{k9}), and (\ref{t10}). Thus $\frac{2}{\kappa}f_{ab}$ is a potential for the conserved current in (\ref{c1}), 
which is the key observation underlying Penrose's quasi-local mass construction, \cite{PR}.

We can write
\be\label{qlm1}
A_{\alpha\beta}Z^\alpha\hat{Z}^\beta:=\frac{2}{\kappa}\oint_{\p\Sigma}f_{ab}d\Sigma^{ab}=\int_\Sigma T_{ab}K^bd\Sigma^a,\ee
where $Z^\alpha,\hat{Z}^\beta$ are the elements of $\mathbb{T}$ corresponding to $\omega^A,\hat{\omega}^B$ respectively, $\p\Sigma$ is the 2-surface bounding $\Sigma$ and $d\Sigma^{ab}$ is the normal 2-form to $\p\Sigma$. This formula serves to define the \emph{angular momentum twistor} $A_{\alpha\beta}$ for $\p\Sigma$.
\item
From the definition, $A_{\alpha\beta}$ is symmetric. It also has a Hermiticity property:
\[A_{\alpha\beta}I^{\beta\gamma}=I_{\alpha\beta}\bar{A}^{\beta\gamma}\]
since if one uses $\hat{Z}^\alpha=I^{\alpha\beta}\bar{Z}_\beta$ then the Killing vector in (\ref{qlm1}) becomes $\tilde{K}^a$ from section 3  which is real.
\item
When the construction is carried through for linear theory in Minkowski space or in anti--de Sitter space one can also show positivity of $A_{\alpha\beta}$ in that
\[A_{\alpha\beta}I^{\beta\gamma}Z^\alpha\bar{Z}_\gamma\geq 0,\]
provided the source $T_{ab}$ satisfies the Dominant Energy Condition (DEC). 
However for linear theory in de Sitter space one does not have this positivity for a reason already discussed: there are no everywhere time-like Killing vectors.\footnote{In this connection see \cite{P} and \cite{ST}.}  
%

\end{enumerate}

{\bf{Acknowledgement}}

This article arose from a desire to provide basic background material for \cite{ST} and I am grateful to Laszlo Szabados for useful discussions and to the Wigner Research Centre for Physics, Budapest for hospitality during the project.

\end{document}